\documentclass[aps,prd,preprint,showpacs]{revtex4}
\usepackage{graphicx}
\usepackage{subfigure}
\usepackage{hyperref}
\usepackage{amsmath}
\begin{document}
\draft
\title{$ZZ\gamma$ and $Z\gamma\gamma$ anomalous couplings in $\gamma p$ collision at the LHC}

\author{A. Senol}\email{asenol@kastamonu.edu.tr}
\affiliation{Kastamonu University, Department of Physics, 37100,
Kastamonu, Turkey} \affiliation{Abant Izzet Baysal University,
Department of Physics, 14280, Bolu, Turkey}
\begin{abstract}
We study the sensitivity of anomalous $ZZ\gamma$ and $Z\gamma\gamma$
vertex couplings $h_3^{\gamma,Z}$ and $h_4^{\gamma,Z}$, which would
be powerful sign of new physics, via the subprocess $\gamma q\to Z
q$ of the main reaction $pp\to p\gamma p\to Z q X$ at the LHC. We
calculated limits on these couplings at 95\% confidence level for
various values of integrated luminosity. It is shown that the $pp\to
p\gamma p\to Z q X$ reaction provides one order of magnitude
improvement in the couplings $h_4^{\gamma,Z}$ compared to the
current experimental limits obtained in events dominated by
$Z\gamma$ production from the LHC and Tevatron.
\end{abstract}
 \maketitle
\section{introduction}
The gauge boson self-interactions are determined by the non-Abelian
$SU(2)_L\times U(1)_Y$ gauge group of the electroweak sector in the
Standard Model (SM). Precision measurements of these interactions
will be important for the test of the SM structure. The tree-level
couplings between the $Z$ boson and the photon ($ZZ\gamma$ and
$Z\gamma\gamma$) vanish in the SM. Any detected signals of these
couplings being from the SM expectations within the experimental
precision would provide crucial clues for new physics beyond the SM.
These new physics effects are parametrized at higher energies via an
effective Lagrangian which reduces to the SM at low energies.

\begin{figure*}[htbp!]
  \includegraphics[width=10cm]{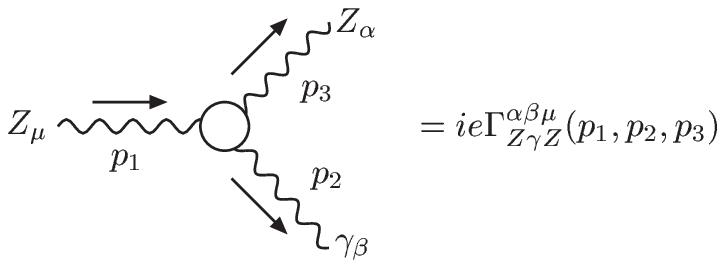}\\
 \caption{Feynman rule for the $ZZ\gamma$ vertex.}\label{vertex}
\end{figure*}

The most general anomalous trilinear $Z\gamma Z$ vertex function,
being consistent with Lorentz and $U(1)_{em}$ gauge invariance, is
given by \cite{Hagiwara:1986vm,Baur:1992cd}:
\begin{eqnarray}\label{eq1}
\Gamma_{Z\gamma Z}^{\alpha\beta\mu}(p_{1},p_{2},p_{3})
&=&\frac{p_{3}^{2}-p_{1}^{2}}{m_{Z}^{2}}
\bigg[h_1^Z(p_2^{\mu}g^{\alpha\beta}-p_2^{\alpha}g^{\mu\beta})+\frac{h_2^Z}{M_Z^2}p_3^{\alpha}\left[(p_3\cdot p_2)g^{\mu\beta}-p_2^{\mu}p_3^{\beta}\right]\nonumber\\
&&+h_{3}^{Z}
\epsilon^{\mu\alpha\beta\rho}p_{2\rho}+\frac{h_{4}^{Z}}{m_{Z}^{2}}
p_{3}^{\alpha}\epsilon^{\mu\beta\rho\sigma}p_{3\rho}p_{2\sigma}
\bigg]
\end{eqnarray}
where $m_{Z}$ denotes the Z-boson mass. Formalism of this vertex is
depicted in Fig.\ref{vertex} where $e$ is the charge of the proton.
Further, the photon and $Z$ boson in the final state are on-shell
while the $Z$ boson in the initial state is off-shell.
The most general $Z\gamma\gamma$ vertex function can be obtained
from Eq.(\ref{eq1}) with the replacements:
\begin{eqnarray}
\frac{p_{3}^{2}-p_{1}^{2}}{m_{Z}^{2}}\to \frac{p_{3}^{2}}{m_{Z}^{2}}
,\;\;\; h_{i}^{Z}\to h_{i}^{\gamma} ,\;\;\; i=1,...,4
\end{eqnarray}
Here the overall factor $p_3^2$ in the $Z\gamma\gamma$ vertex
function results in electromagnetic gauge invariance, while the
factor $p_3^2-p_1^2$ in the $Z\gamma Z$ vertex function
(Eq.(\ref{eq1})) ensures Bose symmetry.

The $h_{i}^{Z}$ and $h_{i}^{\gamma}$ coupling constants in
Eq.(\ref{eq1}) have to be described by means of the energy-dependent
form factors in a dipolelike form due to the restriction of the
$ZZ\gamma$ and $Z\gamma\gamma$ couplings to their SM values at high
energies at tree-level unitary
\cite{Cornwall:1973tb,Cornwall:1974km,Llewellyn Smith:1973ey}.
Following Ref. \cite{Baur:1992cd}, the generalized dipolelike form
factors are described:
\begin{eqnarray}
h_{i}^{V}(\hat{s})= \frac{h_{i0}^{V}}{(1+\hat{s}/\Lambda^{2})^{3}}
~~;~~i=1,3 \\
h_{i}^{V}(\hat{s})= \frac{h_{i0}^{V}}{(1+\hat{s}/\Lambda^{2})^{4}}
~~;~~i=2,4
\end{eqnarray}
$h_{3,4}^V (h_{1,2}^V)$ couplings are CP-conserving (CP-violating).
All the $h_i^V$ couplings vanish at the tree-level in the SM. The
CP-violating couplings always cause completely imaginary amplitudes
that do not interfere with amplitudes of SM diagrams; thus, we are
interested in the CP-conserving couplings. Also, we assume that the
new physics scale $\Lambda$ is above the collision energy
$\sqrt{\hat{s}}$ to neglect the energy dependence of the form
factors.

The 95 \% C.L. intervals for anomalous $ZZ\gamma$ and
$Z\gamma\gamma$ couplings have been provided by ATLAS
\cite{Aad:2012mr} for an integrated luminosity ($L_{int}$) of 1.02
fb$^{-1}$ and $\Lambda=\infty$ , CMS \cite{Chatrchyan:2011rr} for
$L_{int}$=36 pb$^{-1}$ and $\Lambda=\infty$, D0 \cite{Abazov:2011qp}
for $L_{int}$=7.2 fb$^{-1}$ and $\Lambda=\infty$, CDF
\cite{Aaltonen:2011zc} for $L_{int}$=5.1 fb$^{-1}$ and $\Lambda=1.5$
TeV and LEP \cite{Alcaraz:2006mx} obtained from $Z\gamma$ events
which are given in Table \ref{limits}.
\begin{table}
\caption{Summary table of limits at the 95\% C.L. on anomalous
$ZZ\gamma$ and $Z\gamma\gamma$ couplings from $Z\gamma$
events.\label{limits}}
 \small{\begin{tabular}{lccccc}
  \hline
  Parameters & ATLAS                   & CMS                     & D0                     & CDF                      &LEP
  \\\hline

  $h_3^{\gamma}$ & (-0.028,0.027)     & (-0.07, 0.07)     & (-0.027, -0.027)  & (-0.022, 0.020) &(-0.049,0.008) \\
  $h_3^Z$        & (-0.022,0.026)     & (-0.05, 0.06)     & (-0.026, 0.026)   & (-0.020, 0.021) &(-0.20,0.07) \\
  $h_4^{\gamma}$ & (-0.00021,0.00021) & (-0.0005, 0.0006) & (-0.0014, 0.0014) & (-0.0008,0.0008)&(-0.002,0.034) \\
  $h_4^Z$        & (-0.00022,0.00021) & (-0.0005, 0.0005) & (-0.0013, 0.0013) & (-0.0009,0.0009)&(-0.05,0.12) \\
  \hline
\end{tabular}}
\end{table}

Probing on $ZZ\gamma$ and $Z\gamma\gamma$ couplings has been studied
in the $pp$
\cite{Baur:1992cd,Baur:1997kz,Baur:2000ae,Choudhury:2000bw},
$e^+e^-$
\cite{Rizzo:1996ge,Walsh:1998qt,Walsh:2002gm,Atag:2003wm,Atag:2004cn,GutierrezRodriguez:2008tb,Ananthanarayan:2011fr},
and $ep$ \cite{Coutinho:2001ki,TurkCakir:2009zz} colliders. In this
work, we focus on limits of the anomalous $h_3^V$ and $h_4^V$
couplings via the subprocess $\gamma q\to Z q$ of the main reaction
$pp\to p\gamma p\to Z q X$ at the LHC. Here, the quasireal photons
emitted from one proton beam are described by equivalent photon
approximation (EPA) \cite{Budnev:1974de,Ginzburg:1981vm} and can
interact with quarks coming from the other proton beam. Any process
in a $\gamma$-proton collision is different from the pure deep
inelastic scattering process as a result of two distinctive
experimental features. Namely, the first feature is the quasireal
photons emitted from the proton have a low virtuality and are
scattered with small angles from the beam pipe in the framework of
EPA, and for this reason photon-emitting intact protons get away
from the central detector without being detected. This leads to a
reduction in the energy deposit in the corresponding forward region.
Therefore, one of the forward regions of the central detector has a
considerable lack of energy, i.e. forward rapidity gaps. Applying a
selected cut on this quantity, ordinary $pp$ deep inelastic
processes can be sorted out. Another feature is provided by forward
detectors. Particles with large pseudorapidity can be detected from
forward detectors. If the intact proton emitting a photon is
scattered with a large pseudorapidity, it escapes from the central
detectors. These protons leave a characteristic sign in the forward
detectors for $\gamma$-proton collision. These features increase
interest in probing new physics via photon-induced processes at the
LHC in the literature
\cite{Khoze:2001xm,Kepka:2008yx,deFavereaudeJeneret:2009db,Albrow:2010yb,Sahin:2011yv,Gupta:2011be,Sahin:2012zm,Sahin:2012mz,Sahin:2012ry}.

\section{The cross sections of the subprocess $\gamma q\to Z q$  }
\begin{figure*}[htbp!]
  \includegraphics[width=16cm]{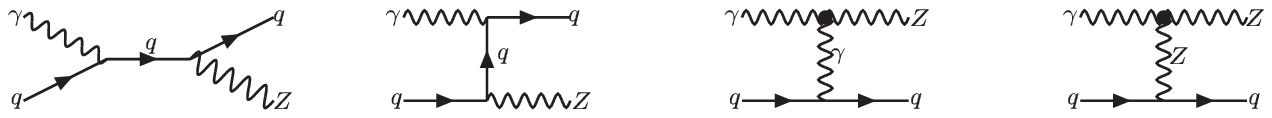}\\
 \caption{Tree-level Feynman diagrams for the subprocess $\gamma q\to Z q $ ($q=u,\bar{u},d,\bar{d},b,\bar{b},s,\bar{s},c,\bar{c}$).}\label{fd}
\end{figure*}

The subprocess $\gamma q\to Z q$ of the main reaction $pp\to p\gamma
p\to Z q X$ at the tree level receives contributions from four
Feynman diagrams, as shown in Fig. \ref{fd}. The last two diagrams
account for the anomalous $Z\gamma\gamma$ and $ZZ\gamma$ couplings,
and the others depict the SM contributions. The total cross section
for the subprocess $\gamma q\to Z q$ is obtained by integrating the
cross sections over the photon and quark distributions, where
$q=u,\bar{u},d,\bar{d},b,\bar{b},s,\bar{s},c,\bar{c}$. All
calculations were performed by means of the computer package CalcHEP
\cite{Belyaev:2012qa}, after implementation of the vertex functions
Eq. (\ref{eq1}). During calculations, we use parton distribution
functions library CTEQ6L \cite{Pumplin:2002vw} and the photon
spectrum in the EPA \cite{Budnev:1974de} embedded in CalcHEP.

The photon spectrum in EPA as a function of photon energy
$E_{\gamma}$ and its virtuality $Q^2$ is given by the following
formula \cite{Budnev:1974de,Sahin:2012mz,Piotrzkowski:2000rx}:
\begin{eqnarray}
\label{spectrum1}
\frac{dN_\gamma}{dE_{\gamma}dQ^{2}}=\frac{\alpha}{\pi}\frac{1}{E_{\gamma}Q^{2}}
[(1-\frac{E_{\gamma}}{E})
(1-\frac{Q^{2}_{min}}{Q^{2}})F_{E}+\frac{E^{2}_{\gamma}}{2E^{2}}F_{M}]
\end{eqnarray}
where $\alpha$ is the fine structure constant, and $Q^{2}_{min}$
standing for the minimum photon virtuality is given by
\begin{eqnarray*}
&&Q^{2}_{min}=\frac{m^{2}_{p}E^{2}_{\gamma}}{E(E-E_{\gamma})}.
\end{eqnarray*}
Here, $m_p$ is the mass of the proton and $E$ denotes the energy of
the incoming proton beam. The functions of the electric and magnetic
form factors $F_{E}$ and $F_{M}$ are displayed by
\begin{eqnarray*}
&&F_{E}=\frac{4m^{2}_{p}G^{2}_{E}+Q^{2}G^{2}_{M}}
{4m^{2}_{p}+Q^{2}},\;\; \;\;\; F_{M}=G^{2}_{M}\\
&&G^{2}_{E}=\frac{G^{2}_{M}}{7.78}=(1+\frac{Q^{2}}{0.71\mbox{GeV}^{2}})^{-4}
\end{eqnarray*}
The cross section of the process $pp\to p\gamma p\to Z q X$ can be
expressed by integrating the cross section for the subprocess
$\gamma q\to Z q$ over the photon and quark spectra
\begin{eqnarray}
\sigma\left(pp\to p\gamma p\to Z q
X\right)=\int_{Q^{2}_{min}}^{Q^{2}_{max}} {dQ^{2}}\int_{x_{1\;
min}}^{x_{1\;max}} {dx_1 }\int_{x_{2\; min}}^{x_{2\;max}} {dx_2}
\left(\frac{dN_\gamma}{dx_1dQ^{2}}\right)\left(\frac{dN_q}{dx_2}\right)\hat{\sigma}_{
\gamma q\to Z q}(\hat s)
\end{eqnarray}
where, $x_1=\frac{E_\gamma}{E}$, and $x_2$ is the momentum fraction
of the proton's momentum carried by the quark when
$\frac{dN_q}{dx_2}$ is the quark distribution function of the
proton. We have considered photon virtuality $\langle
Q^2\rangle\approx0.01 GeV^2$, due to the low virtuality of the
emitted photons in the EPA \cite{Piotrzkowski:2000rx}. In our
calculations, we set $Q^2_{max}$=2 GeV$^2$ for which the
contribution to the integral above this value is negligible.

\begin{figure*}[htbp!]
  \includegraphics[width=10cm]{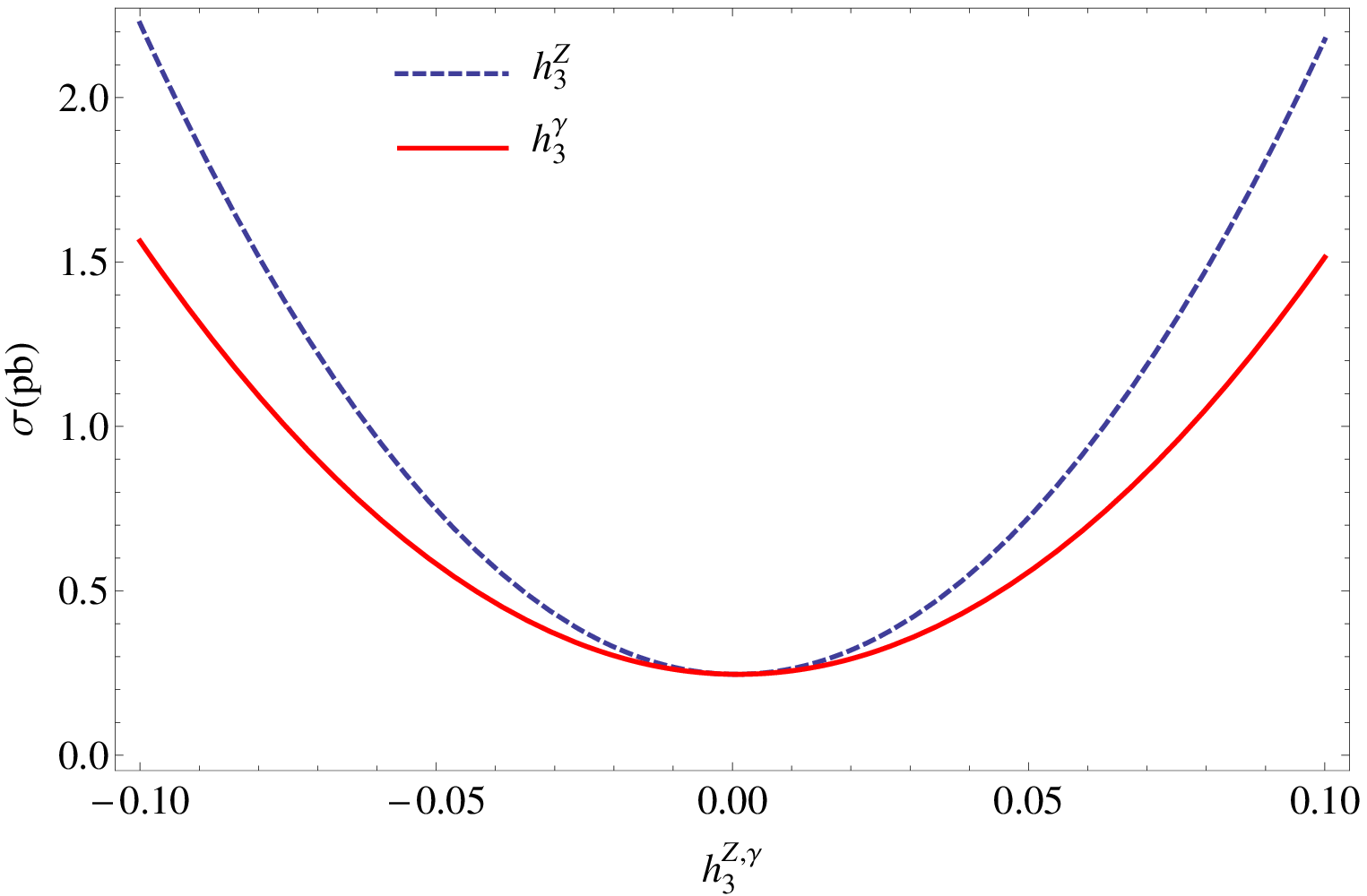}\\
  \caption{The total cross sections depending on anomalous $h_3^{\gamma}$ and $h_3^Z$ couplings for the subprocess $\gamma q\to Z q $
 with taking $\sqrt s$= 14 TeV.}
\label{fig3}
\end{figure*}

\begin{figure*}[htbp!]
  \includegraphics[width=10cm]{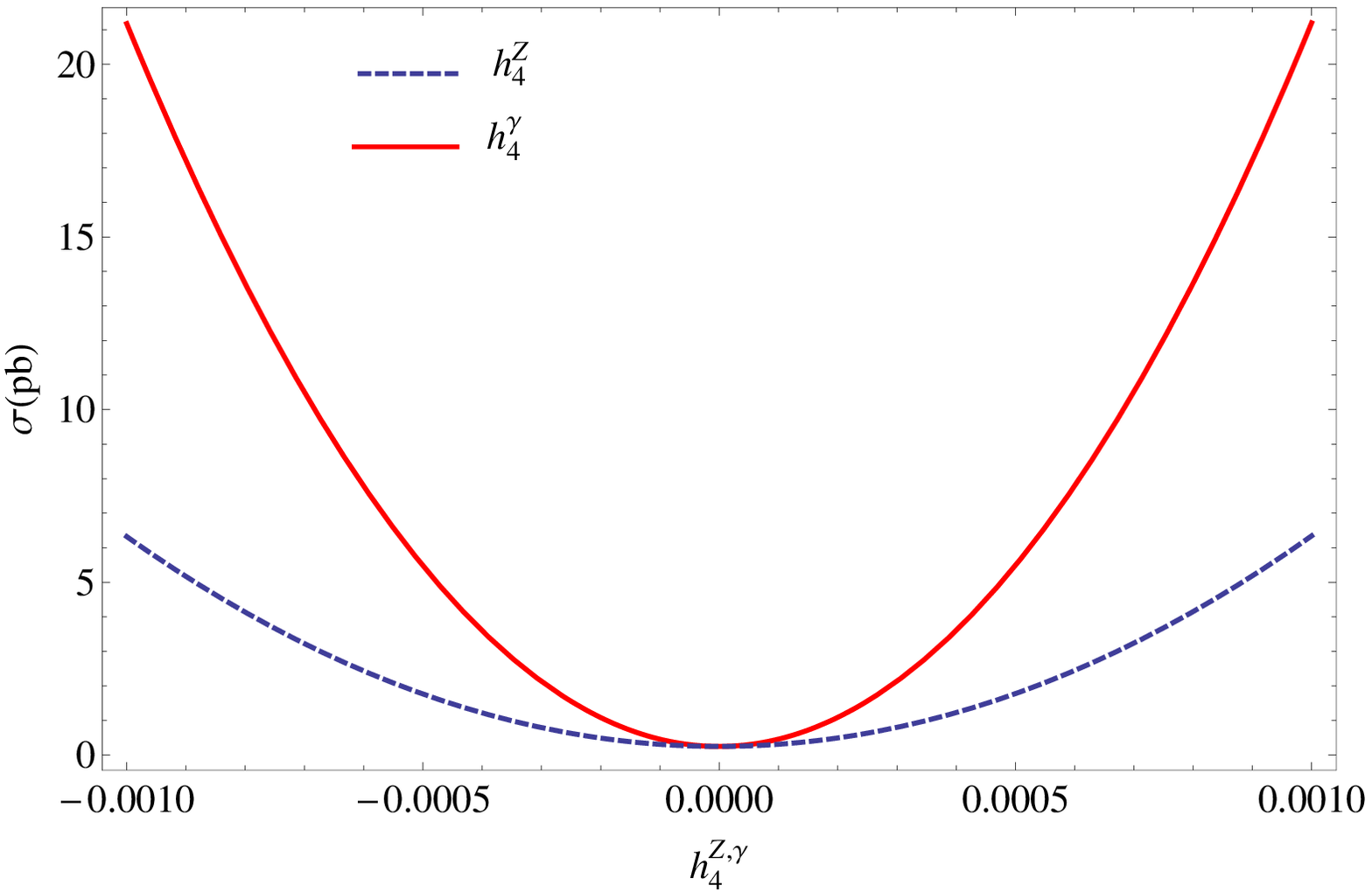}\\
 \caption{The total cross sections depending on anomalous $h_4^{\gamma}$ and $h_4^Z$ couplings for the subprocess $\gamma q\to Z q $
 with taking $\sqrt s$= 14 TeV.}\label{fig4}
\end{figure*}

In Figs. \ref{fig3} and \ref{fig4}, we plot the total cross section
of the subprocess $\gamma q\to Z q $ as a function of anomalous
$h_3^{\gamma, Z}$ and $h_4^{\gamma,Z}$ couplings at the center of
mass energy of 14 TeV. In these figures, only one of the anomalous
couplings is kept to be different from zero. As seen from the
figures, cross sections for $h_3^Z$ couplings are larger as compared
to $h_3^{\gamma}$. In contrast, the cross sections for
$h_4^{\gamma}$ couplings are larger than those of  $h_4^Z$
couplings. This is related to the fact that, the dependencies of the
terms of $h_3^Z$ ($h_4^Z$) and $h_3^{\gamma} (h_4^{\gamma})$ on the
matrix element squared are not the same, because of the presence of
the different overall factors in the vertex functions.
\begin{table}
\caption{One-dimensional limits on $ZZ\gamma$ and $Z\gamma\gamma$
coupling parameters at 95\% C.L. for the subprocess $\gamma q\to Z q
$
 with taking $\sqrt s$= 14 TeV.\label{1D}  }
\begin{tabular}{ccccc}
  \hline
  L(fb$^{-1}$) & $h_3^Z$ & $h_4^Z$ & $h_3^{\gamma}$ & $h_4^{\gamma}$
  \\\hline
  30 & (-0.016, 0.018) & (-0.000099, 0.000098) & (-0.019, 0.022) & (-0.000053, 0.000053) \\
  50 & (-0.014, 0.016) & (-0.000088, 0.000086) & (-0.017, 0.020) & (-0.000047, 0.000046) \\
  100 & (-0.012, 0.013) & (-0.000074, 0.000072) & (-0.014, 0.017) & (-0.000039, 0.000039) \\
  200 & (-0.009, 0.011) & (-0.000062, 0.000061) & (-0.012, 0.014) & (-0.000033, 0.000033) \\
  \hline
\end{tabular}
\end{table}
\section{Limits on the anomalous ZZ$\gamma$ and Z$\gamma\gamma$ couplings}
One-dimensional and two-dimensional $\chi^2$ tests were applied
without a systematic error to obtain 95\% C.L. on the upper limits
of anomalous $h_3^{\gamma, Z}$ and $h_4^{\gamma,Z}$ couplings. The
$\chi^2$ function is
\begin{eqnarray}
\chi^{2}=\left(\frac{\sigma_{SM}-\sigma_{AN}}{\sigma_{SM} \,\,
\delta}\right)^{2}
\end{eqnarray}
where $\delta=\frac{1}{\sqrt{N}}$ is the statistical error. The
number of events are given by $N=S\times E\times\sigma_{SM}\times
L_{int}\times BR(Z\to\l\bar l)$ where $S$ is the survival
probability factor, $E$ denotes the jet reconstruction efficiency,
$L_{int}$ is the integrated luminosity and $l=e^-$ or $\mu^-$. When
calculating the number of events we assume $S=0.7$ and $E=0.6$ for
our process, the same as in Ref. \cite{Sahin:2012mz}. Due to the
overwhelming four jet QCD background, $Z$ bosons decaying
hadronically are not considered here. We applied both cuts for the
transverse momentum of final state quarks to be $p_T^{j}> 15$ GeV
and the pseudorapidity of final state quarks to be $|\eta|< 2.5$,
because ATLAS and CMS have central detectors with a pseudorapidity
coverage $|\eta|< 2.5$.

If a lower cut is applied on the transverse momentum of scattered
protons emitting photons in a photoproduction process, such a cut
helps us to discern a photoproduction process deduced from the usual
$pp$ backgrounds, since the transverse momenta of the scattered
protons are typically $p_T\lesssim 1$ GeV \cite{Albrow:2010yb}.
Therefore, the transverse momentum of an outgoing proton to be
$p_T>0.1$ GeV within the photon spectrum is applied.

According to these restrictions, we have calculated
$\sigma_{SM}=0.39$ pb for $\gamma q\to Z q$
($q=u,\bar{u},d,\bar{d},b,\bar{b},s,\bar{s},c,\bar{c}$) at $\sqrt
s$= 14 TeV. In Table \ref{1D}, we present 95 \% C.L. sensitivity
limits on $h_3^{\gamma, Z}$ and $h_4^{\gamma,Z}$ for various
integrated luminosities by varying one coupling at a time.

The background considered above comes from the subprocess $\gamma
q\to Z q$ of which the final state is composed of an admixture of
light quarks and jets, and dileptons originating from $Z\to l^+l^-$.
In the case of $b$-tagging we assume the efficiency of 60\%, and the
miss-tagging factors for c-quarks and light quarks are taken as 10\%
and 1\%, respectively. Taking all these criteria, the background
cross section is diminished by 2.1\%. Then, the sensitivity of our
bounds are spoiled by about a factor of 1.75 . To illustrate, the
bounds on $h_4^{\gamma}$ and $h_4^{Z}$ became (-0.000069,0.000069)
and (-0.00013,0.00013) for $L_{int}$=100 fb$^{-1}$, respectively.
Besides, the other source of backgrounds is the instrumental
background arising from the calorimeter noise. The calorimeter noise
can be prohibited with a suitable cut on the transverse energy of
jets (e.g. $E_T>$ 40 GeV).

\begin{figure*}[htbp!]
  \includegraphics[width=10cm]{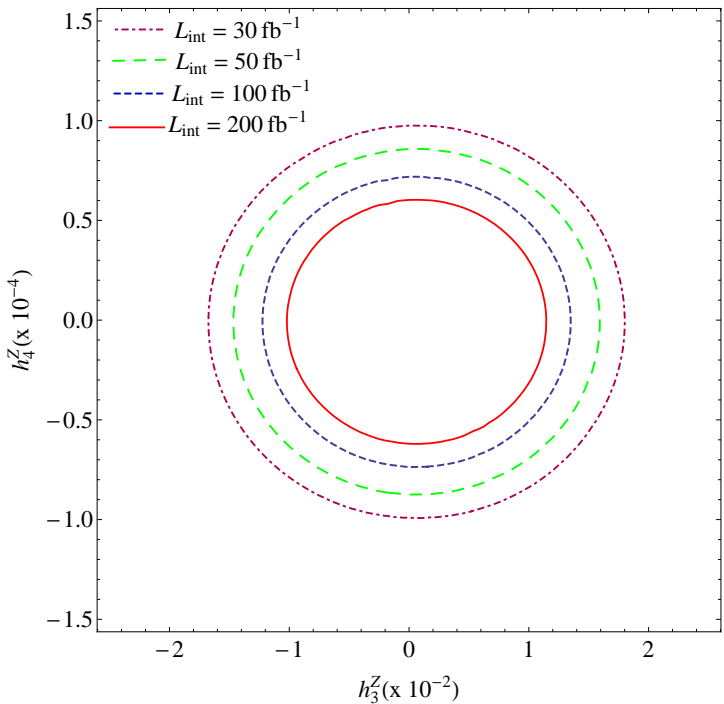}\\
 \caption{Two-dimensional 95\% limit contour for anomalous $h_3^{Z}$ and $h_4^Z$ couplings for the subprocess $\gamma q\to Z q $
 with taking $\sqrt s$= 14 TeV.}\label{fig5}
\end{figure*}

\begin{figure*}[htbp!]
  \includegraphics[width=10cm]{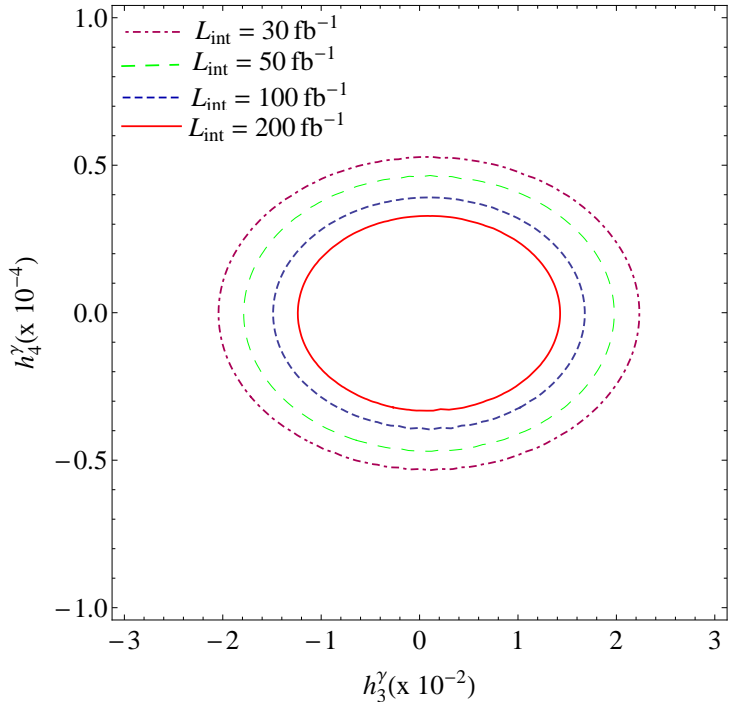}\\
 \caption{Two-dimensional 95\% limit contour for anomalous $h_3^{\gamma}$ and $h_4^{\gamma}$ couplings for the subprocess $\gamma q\to Z q $
 with taking $\sqrt s$= 14 TeV.}\label{fig6}
\end{figure*}

When comparing these limits with the experimental bounds given in
Table \ref{limits}, we can see that the bounds on $h_3^{\gamma, Z}$
in the unitarity violation scheme obtained from ATLAS, D0, and CDF
are of the same order as our bounds, while the $h_4^{\gamma,Z}$
limits are 1 order weaker than our limits. In addition, we show
two-dimensional 95\% C.L. limit contours for $ZZ\gamma$ vertex
couplings $h_3^{ Z}$ and $h_4^{Z}$ in Fig.\ref{fig5} and  for
$Z\gamma\gamma$ vertex couplings $h_3^{ \gamma}$ and $h_4^{\gamma}$
in Fig.\ref{fig6} at $\sqrt s$=14 TeV for various integrated
luminosities. Due to the fact that the $h_4^{Z,\gamma}$ couplings
come from dimension-eight operators, the bounds are more restricted
than those of $h_3^{Z,\gamma}$ which stem from dimension-six.
\section{conclusions}
We have examined the model-independent parametrization of anomalous
$ZZ\gamma$ and $Z\gamma\gamma$ vertex couplings $h_3^{V}$ and
$h_4^{V}$ within the effective operator approach via the subprocess
$\gamma q\to Z q$ of the main reaction $pp\to p\gamma p\to Z q X$ at
the LHC with a center of mass energy of 14 TeV. The potential of the
LHC to probe anomalous $ZZ\gamma$ and $Z\gamma\gamma$ couplings is
analyzed via hadronic $Z\gamma$ production at $\sqrt s$=14 TeV with
the integrated luminosity of 10 and 100 fb$^{-1}$
\cite{Baur:1997kz}. The limits obtained via the $pp\to
Z\gamma+X\to{\not p_T}\gamma+X$ process in Ref.\cite{Baur:1997kz}
are $|h_3^{Z}|<1.9\times10^{-3}(3.4\times 10^{-3})$ and
$|h_4^{Z}|<1.2\times10^{-5}(2.5\times 10^{-5})$ at the LHC with
$L_{int}$=100 (10) fb$^{-1}$. Our results on $h_4^{V}$ are of the
same order with those of Ref.\cite{Baur:1997kz} at $L_{int}$=100
fb$^{-1}$, while the limits on $h_3^{V}$ remains one order lower.
However, a photoproduction process at hadron colliders provides a
rather clean channel compared to the pure deep inelastic process due
to the detection of scattered protons emitting photons by the
forward detectors. Furthermore, the obtained results being related
to the anomalous $ZZ\gamma$ and $Z\gamma\gamma$ vertex couplings
from a photoproduction process are complementary to traditional $pp$
studies. Nevertheless, if we compare the current experimental limits
with the results determined from this work, our limits on the
couplings $h_4^{V}$ with $L_{int}$=30 $fb^{-1}$ are one order better
than the experimental limits obtained from LHC and Tevatron as given
in Table \ref{limits}, while the $h_3^{V}$ couplings are of the same
order as the current experimental limits.
\begin{acknowledgements}
This work is dedicated to my cute daughter, Zeynep Buse Senol. I
would like to thank G. Yildirim and A. T. Tasci for useful
discussions and Orhan Cakir for useful comments.
\end{acknowledgements}
  

\begin{thebibliography}{99}
\bibitem{Hagiwara:1986vm}
  K.~Hagiwara, R.~D.~Peccei, D.~Zeppenfeld and K.~Hikasa,
  %``Probing the Weak Boson Sector in e+ e- ---> W+ W-,''
  Nucl.\ Phys.\ B {\bf 282}, 253 (1987).
  %%CITATION = NUPHA,B282,253;%%
  \bibitem{Baur:1992cd}
  U.~Baur and E.~L.~Berger,
  %``Probing the weak boson sector in $Z \gamma$ production at hadron colliders,''
  Phys.\ Rev.\ D {\bf 47}, 4889 (1993).
  %%CITATION = PHRVA,D47,4889;%%
\bibitem{Cornwall:1973tb}
  J.~M.~Cornwall, D.~N.~Levin and G.~Tiktopoulos,
  %``Uniqueness of spontaneously broken gauge theories,''
  Phys.\ Rev.\ Lett.\  {\bf 30}, 1268 (1973)
  [Erratum-ibid.\  {\bf 31}, 572 (1973)].
  %%CITATION = PRLTA,30,1268;%%
  %294 citations counted in INSPIRE as of 14 Feb 2013
%\cite{}
\bibitem{Cornwall:1974km}
  J.~M.~Cornwall, D.~N.~Levin and G.~Tiktopoulos,
  %``Derivation of Gauge Invariance from High-Energy Unitarity Bounds on the s Matrix,''
  Phys.\ Rev.\ D {\bf 10}, 1145 (1974)
  [Erratum-ibid.\ D {\bf 11}, 972 (1975)].
  %%CITATION = PHRVA,D10,1145;%%
  %915 citations counted in INSPIRE as of 14 Feb 2013
%\cite{}
\bibitem{Llewellyn Smith:1973ey}
  C.~H.~Llewellyn Smith,
  %``High-Energy Behavior and Gauge Symmetry,''
  Phys.\ Lett.\ B {\bf 46}, 233 (1973).
  %%CITATION = PHLTA,B46,233;%%
  %345 citations counted in INSPIRE as of 14 Feb 2013
\bibitem{Aad:2012mr}
  G.~Aad {\it et al.}  [ATLAS Collaboration],
  %``Measurement of $W \gamma$ and $Z \gamma$ production cross sections in $pp$ collisions at $\sqrt{s}=7$ TeV and limits on anomalous triple gauge couplings with the ATLAS detector,''
  Phys.\ Lett.\ B {\bf 717}, 49 (2012)
  [arXiv:1205.2531 [hep-ex]].
  %%CITATION = ARXIV:1205.2531;%%
  %
\bibitem{Chatrchyan:2011rr}
  S.~Chatrchyan {\it et al.}  [CMS Collaboration],
  %``Measurement of $W\gamma$ and $Z\gamma$ production in $pp$ collisions at $\sqrt{s} = 7$ TeV,''
  Phys.\ Lett.\ B {\bf 701}, 535 (2011)
  [arXiv:1105.2758 [hep-ex]].
  %%CITATION = ARXIV:1105.2758;%%
  %
\bibitem{Abazov:2011qp}
  V.~M.~Abazov {\it et al.}  [D0 Collaboration],
  %``$Z\gamma$ production and limits on anomalous $ZZ\gamma$ and $Z\gamma\gamma$ couplings in $p\bar{p}$ collisions at $\sqrt{s}=1.96$ TeV,''
  Phys.\ Rev.\ D {\bf 85}, 052001 (2012)
  [arXiv:1111.3684 [hep-ex]].
  %%CITATION = ARXIV:1111.3684;%%
  %
\bibitem{Aaltonen:2011zc}
  T.~Aaltonen {\it et al.}  [CDF Collaboration],
  %``Limits on Anomalous Trilinear Gauge Couplings in $Z\gamma$ Events from $p\bar{p}$ Collisions at $\sqrt{s} = 1.96$ TeV,''
  Phys.\ Rev.\ Lett.\  {\bf 107}, 051802 (2011)
  [arXiv:1103.2990 [hep-ex]].
  %%CITATION = ARXIV:1103.2990;%%
  %
\bibitem{Alcaraz:2006mx}
  J.~Alcaraz {\it et al.}  [ALEPH and DELPHI and L3 and OPAL and LEP Electroweak Working Group Collaborations],
  %``A Combination of preliminary electroweak measurements and constraints on the standard model,''
  hep-ex/0612034.
 %%CITATION = HEP-EX/0612034;%%Budnev:1974de,Ginzburg:1981vm
%\cite{}
\bibitem{Baur:1997kz}
  U.~Baur, T.~Han and J.~Ohnemus,
  %``QCD corrections and anomalous couplings in $Z \gamma$ production at hadron colliders,''
  Phys.\ Rev.\ D {\bf 57}, 2823 (1998)
  [hep-ph/9710416].
  %%CITATION = HEP-PH/9710416;%%
 \bibitem{Baur:2000ae}
  U.~Baur and D.~L.~Rainwater,
  %``Probing neutral gauge boson selfinteractions in $ZZ$ production at hadron colliders,''
  Phys.\ Rev.\ D {\bf 62}, 113011 (2000)
  [hep-ph/0008063].
  %%CITATION = HEP-PH/0008063;%%
  %\cite{}
\bibitem{Choudhury:2000bw}
  D.~Choudhury, S.~Dutta, S.~Rakshit and S.~Rindani,
  %``Trilinear neutral gauge boson couplings,''
  Int.\ J.\ Mod.\ Phys.\ A {\bf 16}, 4891 (2001)
  [hep-ph/0011205].
  %%CITATION = HEP-PH/0011205;%%
  %
\bibitem{Rizzo:1996ge}
  T.~G.~Rizzo,
  %``$Z$ anomalous couplings and the polarization asymmetry in $\gamma e \to Z e$,''
  Phys.\ Rev.\ D {\bf 54}, 3057 (1996)
  [hep-ph/9602331].
  %%CITATION = HEP-PH/9602331;%%
  %\cite{}
\bibitem{Walsh:1998qt}
  R.~Walsh and A.~J.~Ramalho,
  %``Constraints on the anomalous $Z \gamma \gamma$ and $Z Z \gamma$ vertices at Next Linear Collider energies,''
  Phys.\ Rev.\ D {\bf 57}, 5908 (1998).
  %%CITATION = PHRVA,D57,5908;%%
  %\cite{}
\bibitem{Walsh:2002gm}
  R.~Walsh and A.~J.~Ramalho,
  %``Probing the anomalous z $\gamma \gamma$ and z $\gamma$ z vertices in radiative moller scattering at next linear collider energies,''
  Phys.\ Rev.\ D {\bf 65}, 055011 (2002).
  %%CITATION = PHRVA,D65,055011;%%
%\cite{}
\bibitem{Atag:2003wm}
  S.~Atag and I.~Sahin,
  %``ZZ gamma and Z gamma gamma couplings in gamma e collision with polarized beams,''
  Phys.\ Rev.\ D {\bf 68}, 093014 (2003)
  [hep-ph/0310047].
  %%CITATION = HEP-PH/0310047;%%
  %\cite{}
\bibitem{Atag:2004cn}
  S.~Atag and I.~Sahin,
  %``ZZ gamma and Z gamma gamma couplings at linear e+ e- collider energies with the effects of Z polarization and initial state radiation,''
  Phys.\ Rev.\ D {\bf 70}, 053014 (2004)
  [hep-ph/0408163].
  %%CITATION = HEP-PH/0408163;%%
  %\cite{}
\bibitem{GutierrezRodriguez:2008tb}
  A.~Gutierrez-Rodriguez, M.~A.~Hernandez-Ruiz and M.~A.~Perez,
  %``Probing the ZZgamma and Z gamma gamma Couplings Through the Process e+ e- ---> nu anti-nu gamma,''
  Phys.\ Rev.\ D {\bf 80}, 017301 (2009)
  [arXiv:0808.0945 [hep-ph]].
  %%CITATION = ARXIV:0808.0945;%%
  %\cite{}
\bibitem{Ananthanarayan:2011fr}
  B.~Ananthanarayan, S.~K.~Garg, M.~Patra and S.~D.~Rindani,
  %``Isolating CP-violating $\gamma$ ZZ coupling in $e^+e^- \to \gamma$ Z with transverse beam polarizations,''
  Phys.\ Rev.\ D {\bf 85}, 034006 (2012)
  [arXiv:1104.3645 [hep-ph]].
  %%CITATION = ARXIV:1104.3645;%%
  \bibitem{Coutinho:2001ki}
  Y.~A.~Coutinho, A.~J.~Ramalho, R.~Walsh and S.~Wulck,
  %``Bounds on the $Z \gamma Z$ anomalous couplings from radiative $e p$ scattering at the Very Large Hadron Collider,''
  Phys.\ Rev.\ D {\bf 64}, 115008 (2001).
  %%CITATION = PHRVA,D64,115008;%%

\bibitem{TurkCakir:2009zz}
  I.~Turk Cakir,
  %``Probing anomalous triple gauge boson couplings in gamma p ---> Z b X process,''
  Acta Phys.\ Polon.\ B {\bf 40}, 309 (2009).
  %%CITATION = APPOA,B40,309;%%
  \bibitem{Budnev:1974de}
  V.~M.~Budnev, I.~F.~Ginzburg, G.~V.~Meledin and V.~G.~Serbo,
  %``The Two photon particle production mechanism. Physical problems. Applications. Equivalent photon approximation,''
  Phys.\ Rept.\  {\bf 15}, 181 (1975).
  %%CITATION = PRPLC,15,181;%%
   \bibitem{Ginzburg:1981vm}
  I.~F.~Ginzburg, G.~L.~Kotkin, V.~G.~Serbo and V.~I.~Telnov,
  %``Colliding gamma e and gamma gamma Beams Based on the Single Pass Accelerators (of Vlepp Type),''
  Nucl.\ Instrum.\ Meth.\  {\bf 205}, 47 (1983).
  %%CITATION = NUIMA,205,47;%%
\bibitem{Khoze:2001xm}
  V.~A.~Khoze, A.~D.~Martin and M.~G.~Ryskin,
  %``Prospects for new physics observations in diffractive processes at the LHC and Tevatron,''
  Eur.\ Phys.\ J.\ C {\bf 23}, 311 (2002)
  [hep-ph/0111078].
  %%CITATION = HEP-PH/0111078;%%
%\cite{}
\bibitem{Kepka:2008yx}
  O.~Kepka and C.~Royon,
  %``Anomalous $W W \gamma$ coupling in photon-induced processes using forward detectors at the LHC,''
  Phys.\ Rev.\ D {\bf 78}, 073005 (2008)
  [arXiv:0808.0322 [hep-ph]].
  %%CITATION = ARXIV:0808.0322;%%

    %\cite{}
\bibitem{deFavereaudeJeneret:2009db}
  J.~de Favereau de Jeneret, V.~Lemaitre, Y.~Liu, S.~Ovyn, T.~Pierzchala, K.~Piotrzkowski, X.~Rouby and N.~Schul {\it et al.},
  %``High energy photon interactions at the LHC,''
  arXiv:0908.2020 [hep-ph].
  %%CITATION = ARXIV:0908.2020;%%
  %\cite{}
  %\cite{}
\bibitem{Albrow:2010yb}
  M.~G.~Albrow, T.~D.~Coughlin and J.~R.~Forshaw,
  %``Central Exclusive Particle Production at High Energy Hadron Colliders,''
  Prog.\ Part.\ Nucl.\ Phys.\  {\bf 65}, 149 (2010)
  [arXiv:1006.1289 [hep-ph]].
  %%CITATION = ARXIV:1006.1289;%%
\bibitem{Sahin:2011yv}
  I.~Sahin and A.~A.~Billur,
  %``Anomalous $WW\gamma$ couplings in $\gamma-p$ collision at the LHC,''
  Phys.\ Rev.\ D {\bf 83}, 035011 (2011)
  [arXiv:1101.4998 [hep-ph]].
  %%CITATION = ARXIV:1101.4998;%%
  %\cite{}
\bibitem{Gupta:2011be}
  R.~S.~Gupta,
  %``Probing Quartic Neutral Gauge Boson Couplings using diffractive photon fusion at the LHC,''
  Phys.\ Rev.\ D {\bf 85}, 014006 (2012)
  [arXiv:1111.3354 [hep-ph]].
  %%CITATION = ARXIV:1111.3354;%%
  %\cite{}
\bibitem{Sahin:2012zm}
  I.~Sahin,
  %``Electromagnetic properties of the neutrinos in gamma-proton collision at the LHC,''
  Phys.\ Rev.\ D {\bf 85}, 033002 (2012)
  [arXiv:1201.4364 [hep-ph]].
  %%CITATION = ARXIV:1201.4364;%%
  %\cite{}
\bibitem{Sahin:2012mz}
  I.~Sahin and B.~Sahin,
  %``Anomalous quartic ZZgammagamma couplings in gamma-proton collision at the LHC,''
  Phys.\ Rev.\ D {\bf 86}, 115001 (2012)
  [arXiv:1211.3100 [hep-ph]].
  %%CITATION = ARXIV:1211.3100;%%
  %\cite{}
\bibitem{Sahin:2012ry}
  B.~Sahin and A.~A.~Billur,
  %``Anomalous Wtb couplings in gamma-proton collision at the LHC,''
  Phys.\ Rev.\ D {\bf 86}, 074026 (2012)
  [arXiv:1210.3235 [hep-ph]].
  %%CITATION = ARXIV:1210.3235;%%
 \bibitem{Belyaev:2012qa}
  A.~Belyaev, N.~D.~Christensen and A.~Pukhov,
  %``CalcHEP 3.4 for collider physics within and beyond the Standard Model,''
  arXiv:1207.6082 [hep-ph].
  %%CITATION = ARXIV:1207.6082;%%\cite{Boos:2004kh}
  \bibitem{Pumplin:2002vw}
  J.~Pumplin, D.~R.~Stump, J.~Huston, H.~L.~Lai, P.~M.~Nadolsky and W.~K.~Tung,
  %``New generation of parton distributions with uncertainties from global QCD analysis,''
  JHEP {\bf 0207}, 012 (2002)
  [hep-ph/0201195].
  \bibitem{Piotrzkowski:2000rx}
  K.~Piotrzkowski,
  %``Tagging two photon production at the CERN LHC,''
  Phys.\ Rev.\ D {\bf 63}, 071502 (2001)
  [hep-ex/0009065].
  %%CITATION = HEP-EX/0009065;%%
  %62 citations counted in INSPIRE as of 14 Feb 2013
\end{thebibliography}
\end{document}